\documentclass[runningheads]{llncs}
\usepackage{amsmath}
\usepackage{graphicx}
\usepackage{amsmath}
\usepackage{soul,color}
\usepackage{tikz}
\usepackage{booktabs}
\usepackage{multirow}
\usepackage{graphicx}
\usetikzlibrary{arrows,decorations.markings}
\usepackage{xcolor,colortbl}
\usepackage{float}
\usepackage{xpatch}
\usepackage{subcaption}
\usepgflibrary{arrows}
\newcommand{\mathbbm}[1]{\text{\usefont{U}{bbm}{m}{n}#1}} 
\usepackage[T1]{fontenc}
\DeclareFontFamily{T1}{calligra}{}
\DeclareFontShape{T1}{calligra}{m}{n}{<->s*[1.44]callig15}{}
\DeclareMathAlphabet\mathcalligra   {T1}{calligra} {m} {n}
\DeclareMathAlphabet\mathzapf       {T1}{pzc} {mb} {it}
\DeclareMathAlphabet\mathchorus     {T1}{qzc} {m} {n}
\DeclareMathAlphabet\mathrsfso      {U}{rsfso}{m}{n}
\usepackage{array}
\newcolumntype{P}[1]{>{\centering\arraybackslash}p{#1}}

\begin{document}
\title{Fast Dynamic Perfusion and Angiography Reconstruction using an end-to-end 3D Convolutional Neural Network }
\titlerunning{Fast Deep Dynamic Perfusion and Angiography Reconstruction}
%
\author{Sahar Yousefi\inst{1,*} \and Lydiane Hirschler\inst{1}\and \\Merlijn van der Plas\inst{1} \and Mohamed S. Elmahdy\inst{1}\and \\Hessam Sokooti \inst{1}\and Matthias Van Osch\inst{1} \and
Marius Staring\inst{1,2} }
\authorrunning{S. Yousefi et al.}
%
\institute{Leiden University Medical Center, Radiology, Leiden, The Netherlands \and
Delft University of Technology, Intelligent Systems Department, Delft, The Netherlands\\
\email{s.yousefi.radi@lumc.nl}
}

\maketitle              
\begin{abstract}
Hadamard time-encoded pseudo-continuous arterial spin labeling (te-pCASL) is a signal-to-noise ratio (SNR)-efficient MRI technique for acquiring dynamic pCASL signals that encodes the temporal information into the labeling according to a Hadamard matrix. In the decoding step, the contribution of each sub-bolus can be isolated resulting in dynamic perfusion scans. When acquiring te-ASL both with and without flow-crushing, the ASL-signal in the arteries can be isolated resulting in 4D-angiographic information. However, obtaining multi-timepoint perfusion and angiographic data requires two acquisitions. In this study, we propose a 3D Dense-Unet convolutional neural network with a multi-level loss function for reconstructing multi-timepoint perfusion and angiographic information from an interleaved $50\%$-sampled crushed and $50\%$-sampled non-crushed data, thereby negating the additional scan time. We present a framework to generate dynamic pCASL training and validation data, based on models of the intravascular and extravascular te-pCASL signals. The proposed network achieved SSIM values of $97.3 \pm 1.1$ and $96.2 \pm 11.1$ respectively for 4D perfusion and angiographic data reconstruction for 313 test data-sets. \end{abstract}

\section{Introduction}
Arterial spin labeling (ASL) is a non-invasive MRI technique which uses magnetically labeled blood water as an endogenous tracer for assessing cerebral blood flow (CBF) \cite{ferlay2015cancer}. Hadamard time-encoded(te)-ASL is a time-efficient approach which provides the possibility to combine the superior SNR of ASL to acquire data at different inflow times to obtain dynamic ASL-data \cite{van2018advances}. When Hadamard te-ASL is done with and without flow-crushing, 4D magnetic resonance angiography (MRA) and arterial input function (AIF) measurements can be obtained next to the perfusion scans \cite{petersen2010quasar}. While this approach improves quantification and enhances information content, it is a factor two slower, since both crushed and non-crushed data need to be acquired. Accelerating te-ASL quantification can be done either by acquiring sub-sampled data in k-space or by reducing the rank of the Hadamard matrix. However, these methods can end up reducing image quality and/or signal-to-noise (SNR) ratio.

In this work, we propose an end-to-end 3D convolutional neural network (CNN) for the reconstruction of multi-timepoint 4D MRA and perfusion scans by using half-sampled crushed as well as half-sampled non-crushed Hadamard te-ASL scans, to maintain image quality and provide accurate CBF quantification. Recently, CNNs have shown outstanding performance in medical imaging \cite{yousefi2018esophageal,elmahdy2019robust,gong2018iterative}. However, very few CNN reconstruction techniques have been proposed in the context of MRA and perfusion reconstruction. In \cite{gong2017boosting} a U-net shape CNN for boosting SNR and resolution of ASL scans has been proposed. Guo et al. proposed a CNN based method for improving 3D perfusion image quality by the combined use of single- and multi-delay pseudo-continuous arterial spin labeling (PCASL) and an anatomical scan \cite{Guo2018Improving}. In Guo’s study, ground truth perfusion maps were obtained by positron emission tomography scans. In \cite{ho2016temporal} a temporal CNN approach was proposed for perfusion parameter estimation in stroke. The proposed CNN takes in the signals of interest (i.e., concentration-time curves and the AIF) to produce estimated perfusion parameter maps including cerebral blood volume (CBV), CBF, time-to-maximum, and mean transit time.

In this work, different from the previous works, we employ CNNs in order to accelerate the simultaneous acquisition of 4D MRA and perfusion measurements. One of the challenging issues is the different properties of the outputs of the proposed CNN since MRA is intrinsically sparse and has much more elongated structures than the smooth perfusion map. We tackle this issue by employing different weighting of the loss functions of these two output-types and by balancing extracted samples during training. The proposed CNN leverages the idea of dense blocks \cite{huang2017densely}, arranging them in a typical U-shape \cite{jegou2017one}. Loop connectivity patterns in dense blocks improve the flow of gradients throughout the network and strengthen feature propagation and feature re-usability \cite{yousefi2018esophageal}. In this investigation, we compare the performance of several loss functions: mean square error (MSE), VGG-16 perceptual loss, structural similarity index (SSIM) and multi-level SSIM (ML-SSIM). The main contributions of our work are:
\begin{itemize}
	\item To the best of our knowledge, we are the first to propose acceleration of the reconstruction of 4D MRA and perfusion images using interleaved sub-sampled crushed and non-crushed Hadamard te-ASL scans. To allow sub-sampling, we employed an end-to-end 3D CNN for decoding.
	\item We employed a framework for generating training and validation 4D MRA and perfusion scans by generalizing the Buxton kinetic model for a Hadamard te-ASL signal. Different from \cite{zhao2015rapid}, we consider the kinetic arterial model to take into account the arterial compartment. 
	\item We propose a CNN with a multi-level loss function and compare the proposed method with several loss functions, i.e. MSE, VGG-16 perceptual loss, and SSIM.
\end{itemize}
\section{Proposed approach}
\subsection{Problem}

Reconstruction of dynamic perfusion scans at $H-1$ time points can be performed by the decoding of crushed te-pCASL scans of a Hadamard matrix of rank $H$ \cite{van2018advances}. Reconstruction of dynamic MRA scans at $H-1$ time points, next to the 4D perfusion data, is performed by the decoding of non-crushed te-pCASL scans of a Hadamard matrix of rank $H$ and subtraction of the perfusion data from that \cite{petersen2010quasar}. This process can be formulated as

\begin{equation}\label{eq:fully_sample_decoding}
M\left(\{I_i^\mathchorus{NC}\}, \{I_i^\mathchorus{C}\}\right)_{i=1}^{H} =\{\mathbbm{P}(t), \mathbbm{A}(t)\}_{t=1}^{H-1},
\end{equation}
in which $M$ is the decoding and subtraction function as described earlier \cite{van2018advances,petersen2010quasar}, $I_i^\mathchorus{NC}$ and $I_i^\mathchorus{C}$ are the acquired scans of the $i^{th}$ row of non-crushed and crushed Hadamard te-pCASL datasets, $\mathbbm{P}$ and $\mathbbm{A}$ denote perfusion and angiography scans respectively. 

As mentioned before, obtaining 4D-MRA and perfusion scans require two acquisitions. To accelerate this process with a factor of two, we propose an end-to-end 3D CNN, $M^{\prime}$, which reconstructs 4D-MRA and perfusion data by using interleaved half sampled crushed and half sampled non-crushed Hadamard te-pCASL scans. Therefore, the problem of reconstruction can be re-defined by

\begin{equation}
M^{\prime}\left(\{I_{2\times i-1}^\mathchorus{NC}, I_{2\times i}^\mathchorus{C}\}\right)_{i=1}^{H/2} =\{\mathbbm{P}(t), \mathbbm{A}(t)\}_{t=1}^{H-1}.
\end{equation}

\subsection{Proposed Network}
\begin{figure}[t!]
\includegraphics[width=\textwidth]{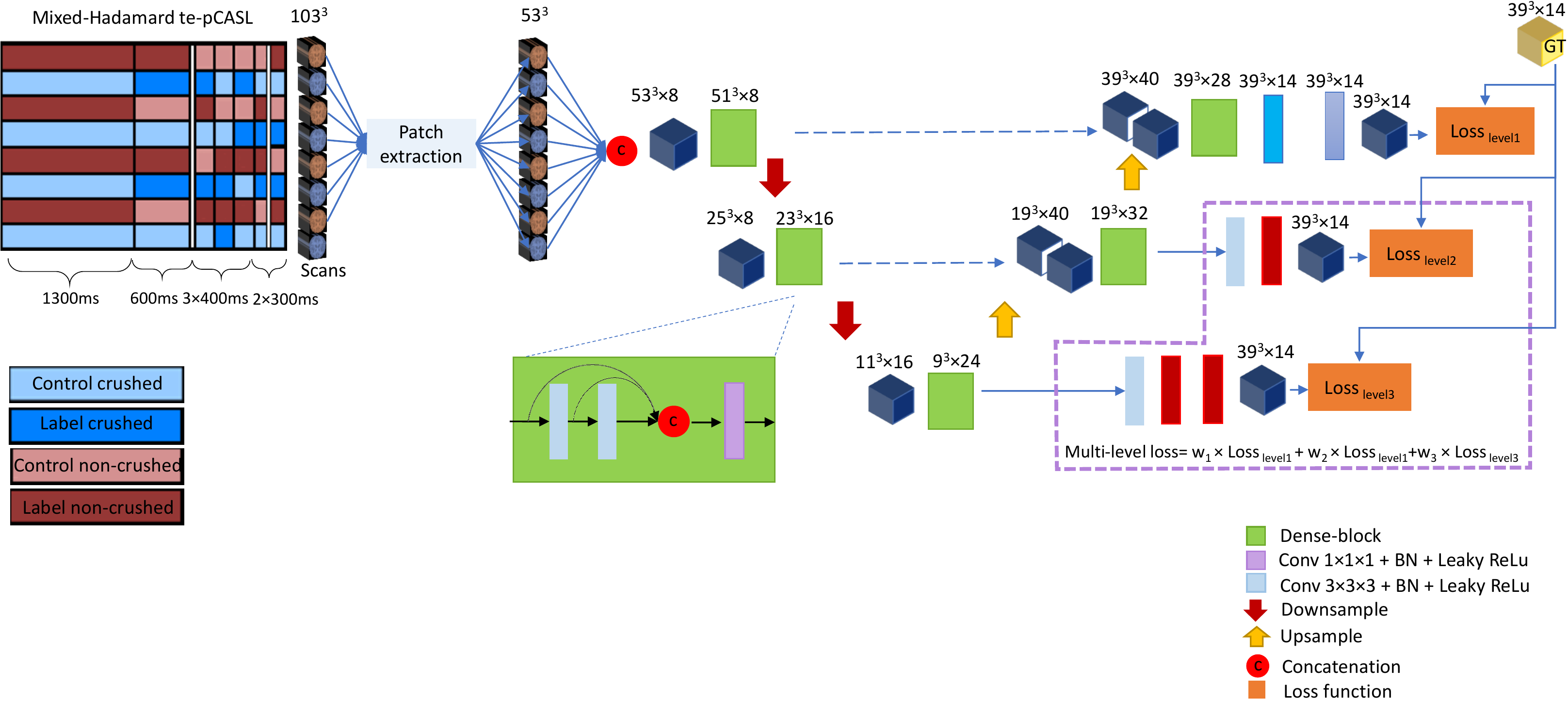}
\caption{Proposed network with single- and multi-level loss functions. The training data contains $50\%$ sub-sampled and interleaved Hadamard-crushed and non-crushed scans. For the single- and multi-level loss functions, $Loss_{level1}$ and $\sum_{i=1}^3 Loss_{level i}$ are considered respectively. GT stands for ground truth, which is the set of angiographic and perfusion data reconstructed by the standard full-sampled decoding approach \cite{van2018advances}.}\label{network}
\end{figure}

Fig. \ref{network} illustrates the proposed network, which takes $50\%$ sub-sampled crushed and $50\%$ sub-sampled non-crushed interleaved ASL data as input and outputs dynamic MRA and perfusion scans. For managing GPU memory, the network was implemented patch-based. The input patches (of size $53^{3}$) are extracted from $50\%$ sub-sampled Hadamard te-crushed and te-non-crushed scans. The outputs of the network are 14 patches of size $39^{3}$ containing perfusion and angiography patches, each at seven different time-points. In this study, we considered a Hadamard matrix of rank 8, so the inputs are 8 patches in total (4 crushed, 4 non-crushed). In each dense block two ($3\times 3\times 3$)conv-BN-leaky ReLu and one ($1\times 1\times 1$)conv-BN-leaky ReLu, as a bottleneck layer, are stacked. Loop connectivity patterns in dense blocks are employed to improve the flow of gradients \cite{huang2017densely}. The bottleneck layers are used to increase the number of feature maps in a tractable fashion, which make the training process easier while leading to a more compact model. A down-sampling unit is followed by one $2\times 2\times 2$ max-pooling layer with a stride of $2\times 2\times 2$. In order to solve the well-known checkerboard issue of the conv-transpose layer, for the up-sampling layer the feature maps are re-sized by a constant trilinear resize convolution kernel, similar to \cite{dong2015image}. In this work we investigate the impact of several loss functions for the defined problem: MSE, which is the $L2$-norm, VGG-16 perceptual loss \cite{johnson2016perceptual}, SSIM which is composed of luminance, contrast and structural error. Later it is shown that among the mentioned loss functions, SSIM has a higher performance in terms of SSIM metric value. Therefore, we propose ML-SSIM, which is calculated based on weighting the SSIM loss function for different levels of the network, see Fig. \ref{network}. 

\subsection{Dataset generation}\label{sec:dataset_generation}

In pCASL the arterial spins are magnetically labeled with a radiofrequency inversion pulse applied below the imaging slices in the neck vessels. The labeled blood then travels via the arteries towards the brain tissue, where they pass from the capillary compartment into the extravascular compartment. After a certain delay time after labeling which is known as the post-labeling delay (PLD) a so-called labeled image is acquired. A control image is acquired without prior labeling and by subtraction of these two images, the perfusion image can be generated. For the Hadamard te-pCASL technique, the labeling module (the typical duration of 3-4 seconds) is divided into several blocks (sub-boli) and a Hadamard matrix is used to determine whether a block will be played-out in label or control condition. For each voxel the Hadamard te-pCASL signal can contain both perfusion signal as well as label still residing in the arteries, i.e. angiography signals.

Since it is difficult to acquire substantial amounts of real data, we propose to model the input data, allowing to generate a sufficient amount of training data. The ground truth output data is created by decoding fully sampled Hadamard te-ASL crushed and non-crushed data \cite{van2018advances}. For this purpose, we create datasets based upon a tracer kinetic model for the Hadamard time-encoded pCASL signal that describes the signal a function of arterial arrival time (AAT), bolus arrival time (BAT) and CBF. In this study, for calculating the signal, the AAT and BAT information are obtained from in vivo data. The CBF maps are taken from the BrainWeb dataset by assigning CBF-values to white matter (WM), gray matter (GM) and cerebrospinal fluid (CSF). 

\begin{figure}[!tb]
\includegraphics[width=12cm]{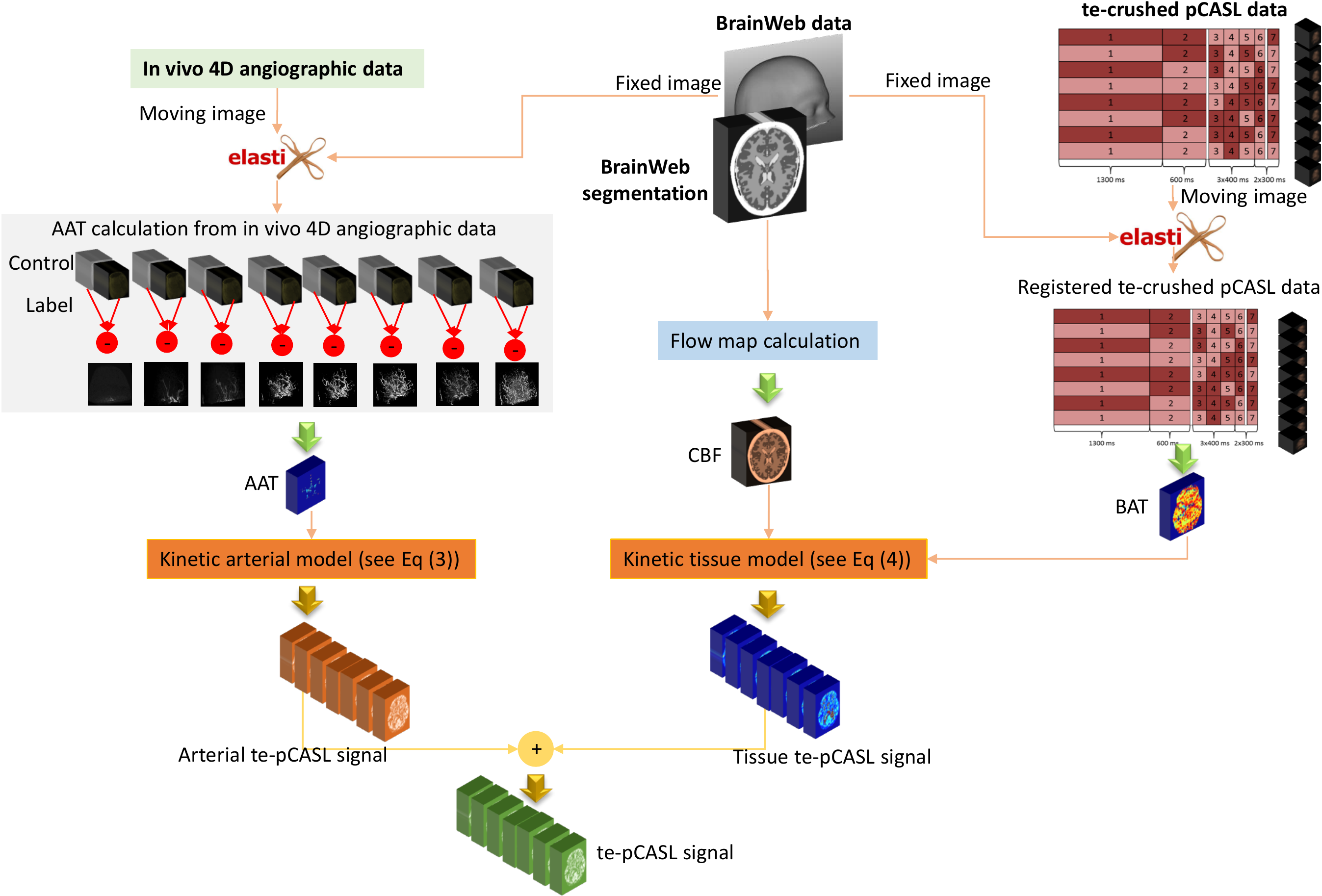}

\caption{Data generator framework for one subject, the inputs of the framework (shown in bold) are: in vivo information (include BAT, AAT, and CBF) and anatomical data from BrainWeb \cite{cocosco1997brainweb}, the outputs are pCASL scans which by decoding the perfusion and angiographic maps are obtained (see Section \ref{sec:dataset_generation}), Left: a map of the arrival time of the label in arteries ('arterial arrival time' or AAT) is created from high resolution 4D MRA ASL scan (Cinema, 8 time-points with 200ms temporal resolution, and a spatial resolution of $0.82 \times 0.82 \times 1.02$ mm) and the data are registered to a subject from BrainWeb dataset, and leads to an AAT map for the subject, then the AAT map is fed into Eq (\ref{eq:arterial_signal}) to calculate the kinetic arterial model. Middle pipeline: using the GM and WM segmentation of the subject and assigning literature values to the flow map (or CBF) of GM and WM the flow map is calculated, which serves as one of the inputs to the kinetic model of the tissue (Eq \ref{eq:tissue_signal}). Right: an in vivo te-crushed pCASL data is registered to the same subject of BrainWeb and by Hadamard decoding the registered data, the arrival of the label at tissue level ('bolus arrival time', or BAT-map) is calculated. This serves as the other input to the kinetic model of the tissue (Eq \ref{eq:tissue_signal}).  The tissue signal and arterial signal are summed together to form the te-pCASL.} \label{datagenerator}
\end{figure}

Fig. \ref{datagenerator} shows the proposed framework for synthetically generating training and validation datasets. For this goal, we leverage the well-known Buxton kinetic model \cite{buxton1998general}, which has been defined for normal ASL, and defined a tracer delivery function (for tissue voxels and arteries) and a tracer accumulation (perfusion) function for each bolus of Hadamard encoded labeling scheme. The final kinetic model is then generated by performing the convolution of the AIF and the residue function. Equations (\ref{eq:arterial_signal}) and (\ref{eq:tissue_signal}) define the obtained model for large arteries and tissue signals for a Hadamard scheme of 8 encoding steps respectively. These equations can be generalized for Hadamard matrices of higher rank. 

\begin{equation}\label{eq:arterial_signal}
 S_{artery} = \begin{cases} 
 0 & \text{if $t<\Delta t_b$}\\
M_{0a}\cdot aCBV \cdot L_r(b)\times e^{\frac{-\Delta t_b}{T_{1b}}} &\text{if $\Delta t_{b} + \sum_{b^\prime=1}^{b-1}\tau_{b^\prime}\leq t<\Delta t_{b} + \sum_{b^\prime=1}^{b}\tau_{b^\prime}$} \\
0 & \text{if $t\geq \Delta t_{b} + \sum_{b^\prime=1}^{N}\tau_{b^{\prime}}$}\\
 \end{cases} 
\end{equation}
\begin{equation}\label{eq:tissue_signal}
S_{tissue}=\begin{cases}
 0 & \text{if $t<\Delta t_b$}\\
 \gamma\Gamma_{\beta=0}& \text{if $\Delta t_a \leq t <\Delta t_a + \tau_1$} \\
 \gamma\left[\Gamma_{\beta=1}+\Xi_{1:1}\right]& \text{if $\Delta t_a+\tau_1 \leq t <\Delta t_a +\sum_{b=1}^2 \tau_b$} \\
 \gamma \left[\Gamma_{\beta=B-1}+\Xi_{B-1:1}\right] &\text{if $\Delta t_a+\sum_{b=1}^{B-1}\tau_b\leq t<\Delta t_a + \sum_{b=1}^B \tau_b; B\in\left[3,7\right]$}\\
 \gamma \Xi_{N:1}& \text{if $t\geq \Delta t_a+\sum_{b=1}^N\tau_b;N=7$}
\end{cases}
\end{equation}
in which $\tau_b$ is label duration for the $b^{th}$ sub-bolus, $N$ is the number of sub-boluses, $M_{0a}$ is the magnetization of arterial blood, $\Delta t_a$ is AAT which represents the arrival time of the labeled blood in the artery, $\Delta t_b$ is BAT which represents the arrival time of labeled blood in the tissue, $T_{1a}$ is the arterial blood relaxation time, $f$ is CBF (millimeter per gram per second), $\kappa$ is static tissue signal, aCBV is arterial cerebral blood volume, 

\begin{equation}
\gamma=M_{0a}\cdot f \cdot e^{\frac{-\Delta t_a}{T_{1a}}} \cdot T_{1a},
\end{equation}
\begin{equation}
\Gamma_\beta=L_r(\beta+1)\left(1-e^{-\frac{t-\Delta t_a-\sum_{b=1}^{\beta}\tau_b}{T_{1a}}}\right),
\end{equation}
and
\begin{equation}
\Xi_{\beta:\beta^\prime}=\sum_{b^\prime=\beta}^{\beta\prime}L_r(b^\prime)\left(e^{-\frac{t-\Delta t_a-\sum_{b=1}^{b^\prime}\tau_b}{T_{1a}}}-e^{-\frac{t-\Delta t_a-\sum_{b=1}^{b^\prime-1}\tau_b}{T_{1a}}}\right).
\end{equation}
$L_r(b)$ is $0$ if the $b^{th}$ bolus in the $r^{th}$ row is control, and it is $1$ if the $b^{th}$ bolus in the $r^{th}$ row is label. For voxels containing large arteries, the pCASL signal can be computed by $S_{voxel} = S_{tissue} + S_{artery}$. The calculated tracer kinetic model is a function of AAT, BAT and CBF. In this study, the anatomic structures are obtained from the BrainWeb database \cite{cocosco1997brainweb}. In order to obtain the tracer signal, the AAT and BAT and blood maps are extracted from in vivo data, then registered with a subject from the BrainWeb dataset by Elastix \cite{klein2010elastix}. The ground truth, i.e. 4D MRA and perfusion scans, are obtained by normal Hadamard decoding of the pCASL images \cite{hirschler2018transit}. To evaluate the generated data, the signal evolution pattern was validated by the Buxton curve model \cite{buxton1998general}.

In this study, the dataset was generated for a Hadamard-8 matrix with seven blocks of respectively 1300, 600, 400, 400, 400, 300 and 300 ms with an additional 265 ms delay before the start of readout. Using the permuted in vivo information (include 6 BAT, 4 AAT) and registering those with the anatomical information from the BrainWeb dataset (consisting of 20 normal subjects and CBF) and calculating the Hadamard te-pCASL (equations (\ref{eq:arterial_signal}) and (\ref{eq:tissue_signal})), this study contains 1564 distinct simulated data-sets each including crushed and non-crushed input data for 8 Hadamard-encodings. By decoding each of the generated crushed and non-crushed te-pCASL data, the corresponding angiographic and perfusion output data at 7-time points, as the ground truth, are obtained. The scans were divided into 1096 subjects for training, 155 for validation and 313 for testing.

\section{Experimental Results}
We implemented the proposed networks in Google's Tensorflow. The patch extraction was done parallel and randomly using a multi-threaded daemon process on the CPU and then patches were fed to the network on the GPU during the training process. To tackle the sparsity of MRA with respect to the perfusion scans, 75 percent of the patches were extracted from the region containing arteries. The input patches were augmented by white noise extracted from a Gaussian distribution with zero mean and random standard deviation between 0 and 5, left-to-right flipping, and random rotation (up to $\pm 18^\circ$).

Evaluation of the proposed networks has been performed by calculating SSIM, MSE, SNR, and peak signal to noise ratio (pSNR), comparing the ground truth reconstruction using full sampling with that of the neural network using $50\%$ subsampling. Table \ref{table:comparison} tabulates a quantitative comparison between the mentioned loss functions. A statistically significant difference (with $p < 0.05$) between ML-SSIM and all the other methods, for perfusion and angiography, can be observed. 

 Fig. \ref{fig:boxplots} depicts the boxplots of the metrics on the test set. The network using the ML-SSIM loss function had a value of $97.3 \pm 1.1$, $6.2\pm 2.4$ and $35.0\pm 3.2$ for SSIM, SNR and pSNR respectively, and the best performance for perfusion reconstruction.  The network with the SSIM loss function had a SSIM of $96.7\pm 12.5$ for angiography reconstruction, i.e. the best performance in terms of SSIM while it does not show a statistically significant difference from the network with the ML-SSIM loss function. Also for angiography reconstruction the network with perceptual loss had the best value for SNR and pSNR while the network with the ML-SSIM loss function with the values of $1.67 \pm 0.52$ and $35.4 \pm 23.2$ for SNR and pSNR respectively had the second rank. 
 
\begin{figure}[!bt]
  \centering
  \begin{center}
  \begin{tikzpicture}
	  \coordinate (A) at (-1,0.4);
  	  \coordinate (B) at ( 5,0.4);
 	  \draw[blue,>=triangle 45, <->] (A) -- (B) node[midway,fill=white] {\emph{Results at multiple PLDs (ms)}};
	 
 	  \coordinate (A) at (6,0.4);
  	  \coordinate (B) at ( 12,0.4);
 	  \draw[green,>=triangle 45, <->] (A) -- (B) node[midway,fill=white] {\emph{Error at multiple PLDs (ms)}};
   \end{tikzpicture}		
  
  \begin{subfigure}[b]{1.3\linewidth}
    \hspace*{-1.5cm}{\put(0,0){\includegraphics[ width=\linewidth]{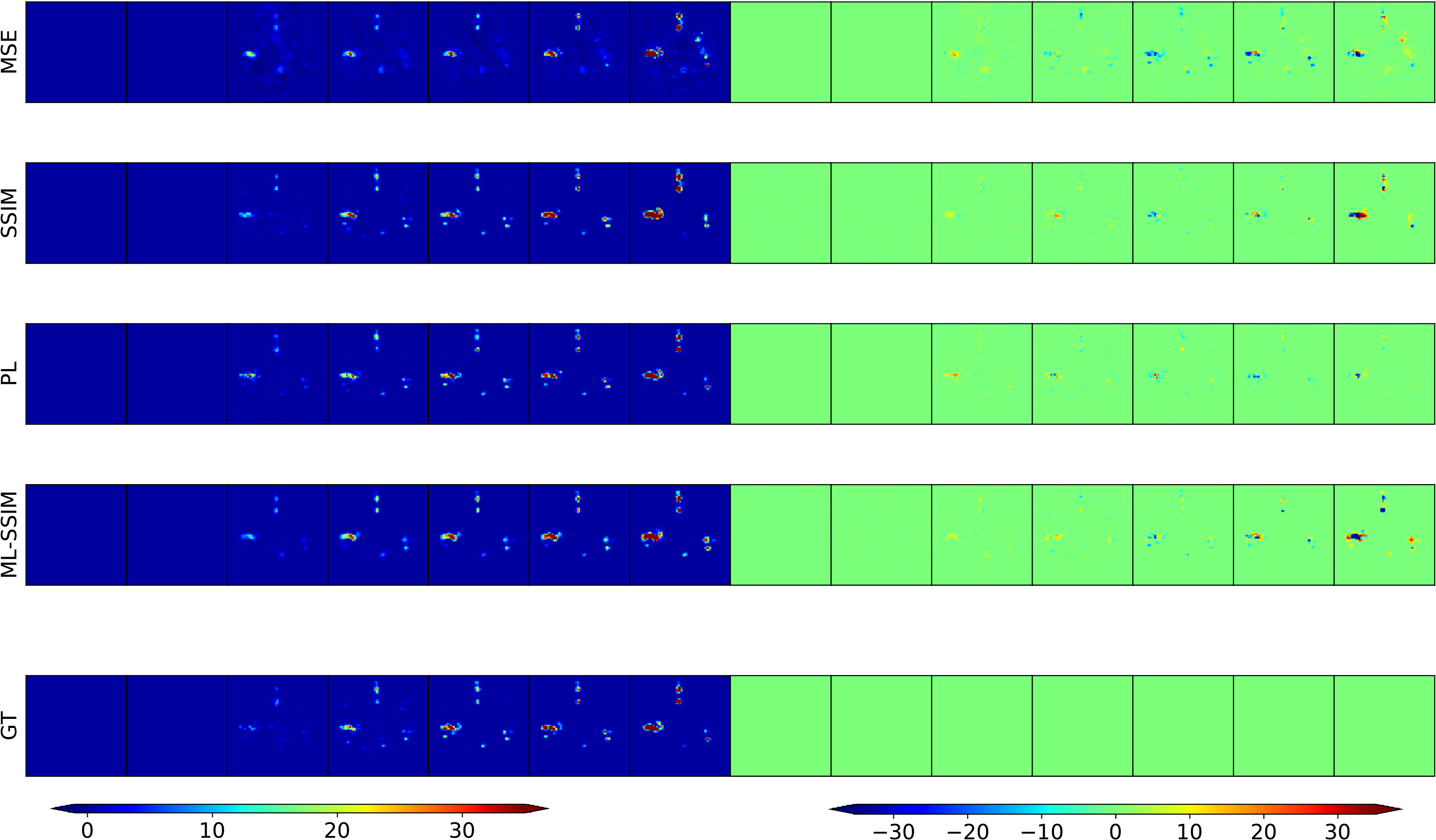}}}
    \put(12,270){265}
    \put(45,270){565}
    \put(77,270){865}
    \put(108,270){1265}
    \put(140,270){1665}
    \put(172,270){2065}
    \put(203,270){2665}          
    \put(237,270){265}
    \put(268,270){565}
    \put(300,270){865}
    \put(330,270){1265}
    \put(360,270){1665}
    \put(395,270){2065}
    \put(428,270){2665}  
    \parbox{6in}{\caption{4D MRA}}
  \end{subfigure}
  
  \begin{subfigure}[b]{1.3\linewidth}
    \hspace*{-1.5cm}{\put(0,0){\includegraphics[width=\linewidth]{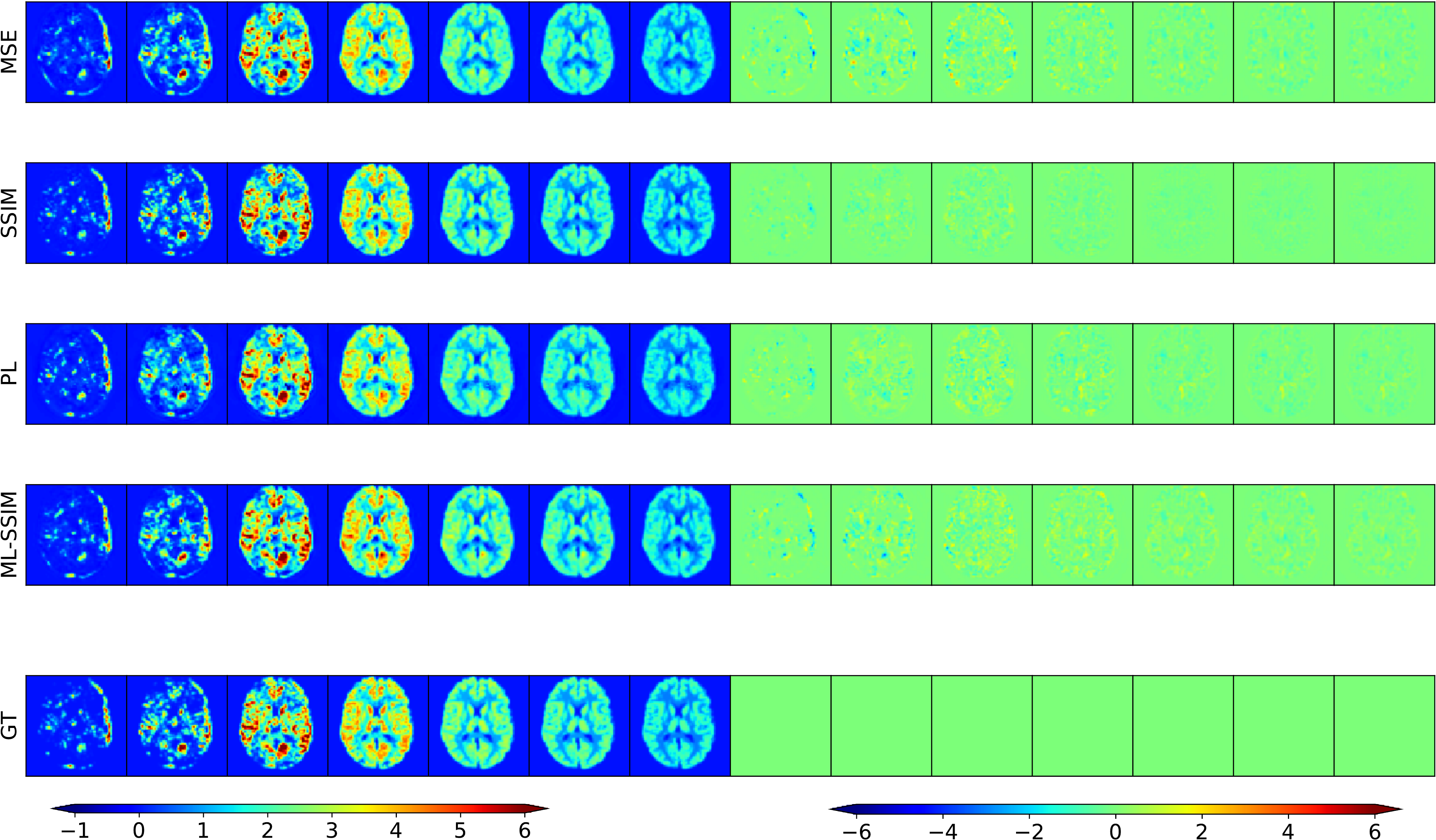}}}
    \parbox{6in}{\caption{4D perfusion}}   
  \end{subfigure}
  
  \end{center}
  \caption{Qualification results, 4D a) MRA and b) perfusion and error at multiple time points after labeling of arterial spins in a single-slice for the different networks, GT stands for ground truth which is obtained from fully sampled decoding and subtracting (see Equation \ref{eq:fully_sample_decoding}).}
  \label{fig:perf_angio}
\end{figure}

Fig. \ref{fig:perf_angio} exemplifies the qualification compassion for 4D MRA and perfusion reconstruction between the different CNNs. The lower SNR and higher pSNR variance for the angiographic data is partially explained by the intrinsic sparsity of that data, especially noticeable in the earlier time points, see Fig. \ref{fig:perf_angio}a.

It takes an average of 205$\pm$232 ms from the ML-SSIM network to reconstruct all perfusion and angiography scans from the interleaved sparsely-sampled crushed and non-crushed data of size $107^3$. 

\begin{table}[!tb]
\small
\centering
\caption{Comparison of the different networks for perfusion and angiographic images (in gray the perfusion and in white the angiographic results), the best results for perfusion and angiography are shown in blue and green respectively, PL stands for perceptual loss. A Wilcoxon signed-rank test is performed between ML-SSIM and other loss functions for perfusion and angiography, where $\dagger$ indicates a statistically significant difference with $p < 0.05$.}
\begin{tabular}{c|c|c|c|c|c|c}
\toprule
\multicolumn{2}{P{3.5cm}|}{Loss function}&SSIM$\%$&MSE&SNR&pSNR&$\#$ of param\\
\hline
  \multirow{4}{*}{MSE} & 
  \cellcolor{gray!25}\multirow{1}{*}{$(\mu\pm\sigma)$} &
   \cellcolor{gray!25}$97.0\pm 1.1 ^{\dagger}$ &
   \cellcolor{gray!25} {$\color{blue}0.02\pm 0.02^{\dagger}$}  &
   \cellcolor{gray!25} $3.84\pm 1.41^{\dagger}$ &
   \cellcolor{gray!25} $30.8\pm 2.3^{\dagger}$ \\   
  & \cellcolor{gray!25}\multirow{1}{*}{$med$} &
   \cellcolor{gray!25}\multirow{1}{*}{96.8} & 
   \cellcolor{gray!25}{0.02 }
   &\cellcolor{gray!25}{4.37}&
   \cellcolor{gray!25}{30.7}&
   \multirow{2}{*}{169,152}\\
  \cline{2-6}
   & \multirow{1}{*}{$(\mu\pm\sigma)$} &
    $70.0\pm 45.5^{\dagger}$ &
     $0.45\pm 2.31^{\dagger}$  & 
     $1.28\pm 0.24^{\dagger}$ & 
     $33.8\pm 22.0^{\dagger}$ \\   
  & \multirow{1}{*}{$med$} & 
  \multirow{1}{*}{99.5} & 
  0.17 &
  1.31&
  46.9&
  \multirow{1}{*}{}\\
\specialrule{.2em}{.1em}{.1em} 
  \multirow{4}{*}{SSIM} &
  \cellcolor{gray!25}\multirow{1}{*}{$(\mu\pm\sigma)$} &
  \cellcolor{gray!25}{$96.3 \pm 1.7  ^{\dagger}$}&
  \cellcolor{gray!25} $0.04\pm 0.04^{\dagger} $&
  \cellcolor{gray!25}{$2.97 \pm 1.26 ^{\dagger}$} &\cellcolor{gray!25}{$28.9 \pm 1.6  ^{\dagger}$} \\   
  & \cellcolor{gray!25}\multirow{1}{*}{$med$} & \cellcolor{gray!25}\multirow{1}{*}{96.99 } & \cellcolor{gray!25}\multirow{1}{*}{0.03  } &\cellcolor{gray!25}\multirow{1}{*}{3.49  }&\cellcolor{gray!25}\multirow{1}{*}{28.44 }&\multirow{2}{*}{169,152}\\
  \cline{2-6}
  & \multirow{1}{*}{$(\mu\pm\sigma)$} &
  {\color{green}$96.7\pm 12.5 $}&
   $0.37\pm 2.64 ^{\dagger}$&
   $1.15 \pm 0.70 ^{\dagger}$ &
   {$33.8 \pm 22.1 ^{\dagger}$ }\\   
  & \multirow{1}{*}{$med$} & 
  \multirow{1}{*}{98.9 } & 
  \multirow{1}{*}{0.20} &
  \multirow{1}{*}{1.12 }&
  \multirow{1}{*}{46.28 }&
  \multirow{1}{*}{}\\
\specialrule{.2em}{.1em}{.1em} 
  \multirow{4}{*}{PL} &
   \cellcolor{gray!25}\multirow{1}{*}{$(\mu\pm\sigma)$} &
   \cellcolor{gray!25}$96.1\pm 1.8^{\dagger}$ &
   \cellcolor{gray!25} $0.05 \pm 0.39^{\dagger}$ &\cellcolor{gray!25}$5.14\pm 1.81^{\dagger}$&
   \cellcolor{gray!25}{$33.7\pm 3.6^{\dagger}$} \\   
  &
   \cellcolor{gray!25}\multirow{1}{*}{$med$} & \cellcolor{gray!25}\multirow{1}{*}{96.29 } &\cellcolor{gray!25} 0.01 &
   \cellcolor{gray!25}6.00 &
   \cellcolor{gray!25}33.19 &
   \multirow{2}{*}{169,152} \\
  \cline{2-6}  
   & \multirow{1}{*}{$(\mu\pm\sigma)$} &
   $ 96.3\pm 12.8 ^{\dagger}$ &
    {\color{green}$0.16 \pm 2.23^{\dagger} $} &
    {\color{green}$3.03\pm 3.21^{\dagger}$}&
    {\color{green}$38.4\pm 25.3^{\dagger}$} \\   
  & \multirow{1}{*}{$med$} & 
  \multirow{1}{*}{99.8 } & 
  0.04 &
  2.30&
  51.34 &
  \multirow{1}{*}{}\\ 
\specialrule{.2em}{.1em}{.1em} 
  \multirow{4}{*}{ML-SSIM} &
   \cellcolor{gray!25}\multirow{1}{*}{$(\mu\pm\sigma)$} &
  \cellcolor{gray!25}{\color{blue}$97.3 \pm 1.1$} &
  \cellcolor{gray!25}{$0.03\pm 0.15$}&
  \cellcolor{gray!25}{\color{blue}$6.18\pm 2.38$}&
  \cellcolor{gray!25}{\color{blue}$35.0 \pm 3.2$} \\ 
  & \cellcolor{gray!25}\multirow{1}{*}{$med$} & \cellcolor{gray!25}\multirow{1}{*}{97.1} &
  \cellcolor{gray!25}  0.01  &
  \cellcolor{gray!25}7.48 &
  \cellcolor{gray!25}34.63&\multirow{2}{*}{181,692}
  \\\cline{2-6}
   & \multirow{1}{*}{$(\mu\pm\sigma)$} &
   {$96.2\pm 11.1$}&
    {$0.44\pm 3.17 $}&
    {$1.67\pm 0.52$} &
    $35.4 \pm 23.2$ \\   
  & \multirow{1}{*}{$med$} & 
  \multirow{1}{*}{99.7 } &
   \multirow{1}{*}{0.10} &
   \multirow{1}{*}{1.76 }&
   \multirow{1}{*}{49.35 }&
   \multirow{1}{*}{}\\

\bottomrule

\end{tabular}\label{table:comparison}

\end{table}


\begin{figure}[t!]
\noindent\includegraphics[height=9cm]{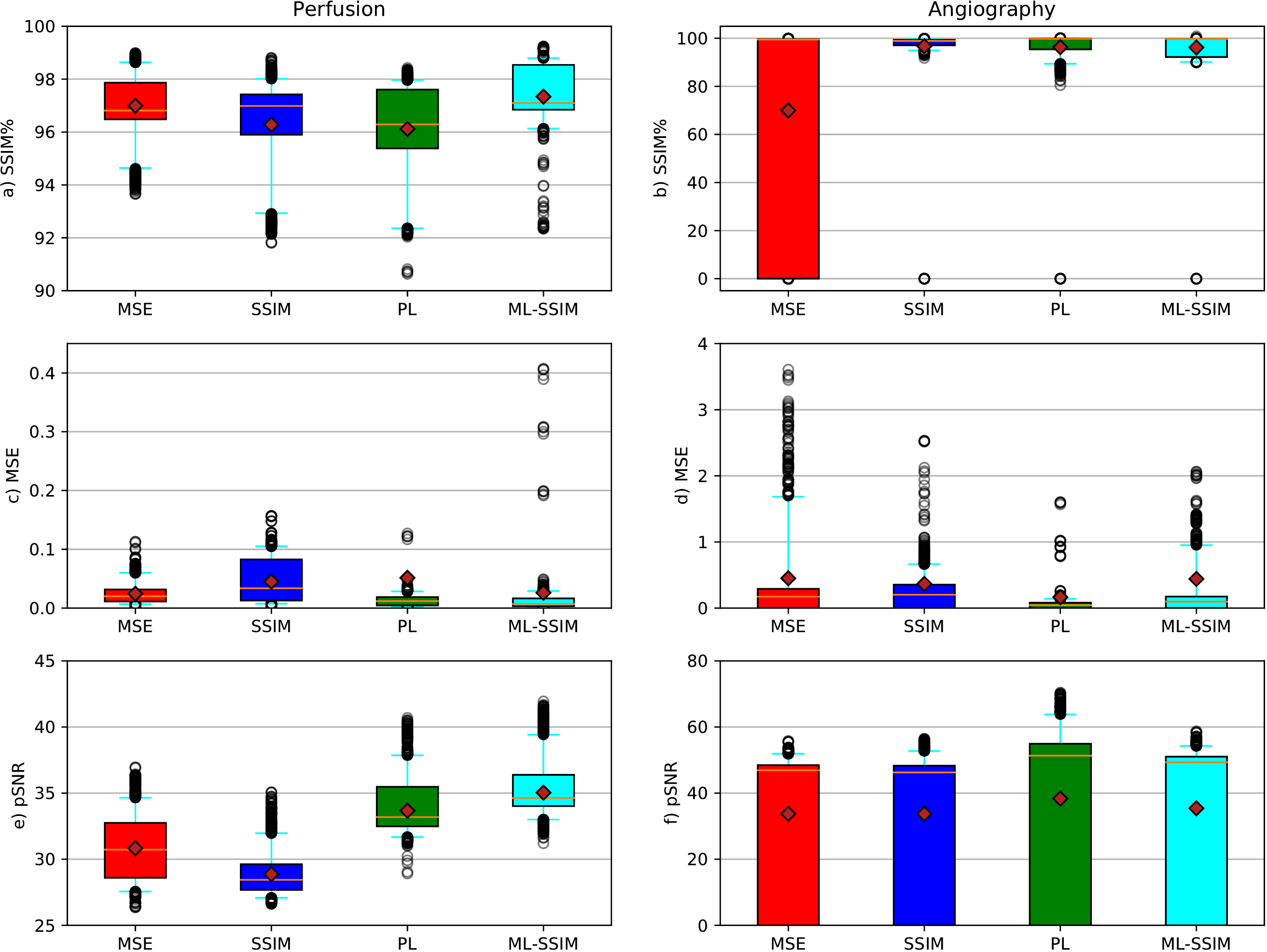} 
\caption{Boxplots for different metrics and CNNs for perfusion and angiographic results, PL stands for VGG-16 perceptual loss. For (e) a few outliers smaller than 25 and for (c)/(d) a few outliers larger than 0.45/4 are not shown.}
\end{figure}\label{fig:boxplots}

\section{Conclusion}
We proposed a 3D end-to-end fully convolutional CNN for accelerating 4D MRA and perfusion reconstruction from half-sampled crushed and non-crushed pCASL data. We leveraged loop connectivity patterns in the network architecture to improve the flow of information during the gradient updates. For training and validation purposes we developed a data generator framework based on the generalized kinetic model for the pCASL signal. The generated dataset included 1096 scans for training, 155 scans for validation and 313 for testing. The proposed network with ML-SSIM loss function achieved a SSIM of ${97.3}\pm{1.1}/{96.2}\pm{11.1}$, MSE of ${0.03}\pm{0.15}/{0.44}\pm{3.17}$, SNR of ${6.18}\pm{2.38}/{1.67}\pm{0.52}$, and pSNR of ${35.0 \pm 3.2}/{35.4 \pm 23.2}$ for perfusion/angiography reconstruction. The lower SNR and higher variance in the pSNR for the angiographic data is partially explained by the intrinsic sparsity of that data, especially noticeable in the earlier time points, see Fig. \ref{fig:perf_angio}a. 


In conclusion, the proposed network obtained promising results for the challenging problem of 4D MRA and perfusion reconstruction. The method, therefore, may assist an accelerated MRI scanning workflow. A further step of this study is enriching the training and validation datasets with in vivo data. \\

\noindent\textbf{Acknowledgements.} This work is financed by the Netherlands Organization for Scientific Research (NWO), VICI project 016.160.351.

\clearpage
%
%
%

\bibliography{references}

\begin{thebibliography}{10}

\bibitem{ferlay2015cancer}
Jacques Ferlay, Isabelle Soerjomataram, Rajesh Dikshit, Sultan Eser, Colin
  Mathers, Marise Rebelo, Donald~Maxwell Parkin, David Forman, and Freddie
  Bray.
\newblock Cancer incidence and mortality worldwide: sources, methods and major
  patterns in globocan 2012.
\newblock {\em International journal of cancer}, 136(5):E359--E386, 2015.

\bibitem{van2018advances}
Matthias~JP van Osch, Wouter~M Teeuwisse, Zhensen Chen, Yuriko Suzuki, Michael
  Helle, and Sophie Schmid.
\newblock Advances in arterial spin labelling mri methods for measuring
  perfusion and collateral flow.
\newblock {\em Journal of Cerebral Blood Flow \& Metabolism}, 38(9):1461--1480,
  2018.

\bibitem{petersen2010quasar}
Esben~Thade Petersen, Kim Mouridsen, and Xavier Golay.
\newblock The quasar reproducibility study, part {II}: Results from a
  multi-center arterial spin labeling test--retest study.
\newblock {\em Neuroimage}, 49(1):104--113, 2010.

\bibitem{yousefi2018esophageal}
Sahar Yousefi, Hessam Sokooti, Mohamed~S Elmahdy, Femke~P Peters, Mohammad
  T~Manzuri Shalmani, Roel~T Zinkstok, and Marius Staring.
\newblock Esophageal gross tumor volume segmentation using a 3d convolutional
  neural network.
\newblock In {\em MICCAI}, pages 343--351. Springer, 2018.

\bibitem{elmahdy2019robust}
Mohamed~S Elmahdy, Thyrza Jagt, R~Th Zinkstok, Yuchuan Qiao, Rahil Shahzad,
  Hessam Sokooti, Sahar Yousefi, Luca Incrocci, CAM Marijnen, Mischa Hoogeman,
  et~al.
\newblock Robust contour propagation using deep learning and image registration
  for online adaptive proton therapy of prostate cancer.
\newblock {\em Medical physics}, 2019.

\bibitem{gong2018iterative}
Kuang Gong, Jiahui Guan, Kyungsang Kim, Xuezhu Zhang, Jaewon Yang, Youngho Seo,
  Georges El~Fakhri, Jinyi Qi, and Quanzheng Li.
\newblock Iterative pet image reconstruction using convolutional neural network
  representation.
\newblock {\em TMI}, 38(3):675--685, 2018.

\bibitem{gong2017boosting}
E~Gong, J~Pauly, and G~Zaharchuk.
\newblock Boosting snr and/or resolution of arterial spin label ({ASL}) imaging
  using multi-contrast approaches with multi-lateral guided filter and deep
  networks.
\newblock In {\em Proceedings of the Annual Meeting of the International
  Society for Magnetic Resonance in Medicine, Honolulu, Hawaii}, 2017.

\bibitem{Guo2018Improving}
Jia Guo, Enhao Gong, Maged Goubran, Audrey~P. Fan, Mohammad~M. Khalighi, and
  Greg Zaharchuk.
\newblock Improving perfusion image quality and quantification accuracy using
  multi-contrast mri and deep convolutional neural networks.
\newblock In {\em ISMRM, Paris, France}, 2018.

\bibitem{ho2016temporal}
King~Chung Ho, Fabien Scalzo, Karthik~V Sarma, Suzie El-Saden, and Corey~W
  Arnold.
\newblock A temporal deep learning approach for {MR} perfusion parameter
  estimation in stroke.
\newblock In {\em 23rd ICPR}, pages 1315--1320. IEEE, 2016.

\bibitem{huang2017densely}
Gao Huang, Zhuang Liu, Laurens Van Der~Maaten, and Kilian~Q Weinberger.
\newblock Densely connected convolutional networks.
\newblock In {\em CVPR}, pages 4700--4708, 2017.

\bibitem{jegou2017one}
Simon J{\'e}gou, Michal Drozdzal, David Vazquez, Adriana Romero, and Yoshua
  Bengio.
\newblock The one hundred layers tiramisu: Fully convolutional densenets for
  semantic segmentation.
\newblock In {\em CVPR}, pages 11--19, 2017.

\bibitem{zhao2015rapid}
Li~Zhao, Samuel~W Fielden, Xue Feng, Max Wintermark, John~P Mugler~III, and
  Craig~H Meyer.
\newblock Rapid 3d dynamic arterial spin labeling with a sparse model-based
  image reconstruction.
\newblock {\em Neuroimage}, 121:205--216, 2015.

\bibitem{dong2015image}
Chao Dong, Chen~Change Loy, Kaiming He, and Xiaoou Tang.
\newblock Image super-resolution using deep convolutional networks.
\newblock {\em TPAMI}, 38(2):295--307, 2015.

\bibitem{johnson2016perceptual}
Justin Johnson, Alexandre Alahi, and Li~Fei-Fei.
\newblock Perceptual losses for real-time style transfer and super-resolution.
\newblock In {\em ECCV}, pages 694--711. Springer, 2016.

\bibitem{buxton1998general}
Richard~B Buxton, Lawrence~R Frank, Eric~C Wong, Bettina Siewert, Steven
  Warach, and Robert~R Edelman.
\newblock A general kinetic model for quantitative perfusion imaging with
  arterial spin labeling.
\newblock {\em MRM}, 40(3):383--396, 1998.

\bibitem{cocosco1997brainweb}
Chris~A Cocosco, Vasken Kollokian, Remi K-S Kwan, G~Bruce Pike, and Alan~C
  Evans.
\newblock Brainweb: Online interface to a 3d mri simulated brain database.
\newblock In {\em NeuroImage}. Citeseer, 1997.

\bibitem{klein2010elastix}
Stefan Klein, Marius Staring, Keelin Murphy, Max~A Viergever, and Josien~PW
  Pluim.
\newblock Elastix: a toolbox for intensity-based medical image registration.
\newblock {\em TMI}, 29(1):196--205, 2010.

\bibitem{hirschler2018transit}
Lydiane Hirschler, Leon~P Munting, Artem Khmelinskii, Wouter~M Teeuwisse, Ernst
  Suidgeest, Jan~M Warnking, Louise van~der Weerd, Emmanuel~L Barbier, and
  Matthias~JP van Osch.
\newblock Transit time mapping in the mouse brain using time-encoded pcasl.
\newblock {\em NMR in Biomedicine}, 31(2):e3855, 2018.

\end{thebibliography}
\bibliographystyle{unsrt}

\end{document}